\documentclass[aps,prd,preprint,12pt,altaffilletter,showpacs]{revtex4}
\usepackage{graphicx}
\usepackage{amssymb}
\usepackage{amsmath} %align environment for multi-line equations
\usepackage{hyperref}
\usepackage{textcomp}
\usepackage{psfrag}
\usepackage{setspace}
\usepackage{url}
\begin{document}
\title{Coherent Emission in Fast Radio Bursts}
\author{J.~I.~Katz$^*$}
\affiliation{Department of Physics and McDonnell Center for the Space
Sciences\\Washington University, St. Louis, Mo. 63130}
\email{katz@wuphys.wustl.edu}
\date{\today}
\begin{abstract}
The fast (ms) radio bursts reported by Lorimer, {\it et al.} \cite{L07} and
by Thornton, {\it et al.} \cite{T13} have extremely high brightness
temperatures if at the inferred cosmological distances.  This implies
coherent emission by ``bunches'' of charges.  FRB, like the giant pulses of
the Crab pulsar, display banded spectra that may be harmonics of plasma
frequency emission by plasma turbulence, and are inconsistent with emission
by charge distributions moving relativistically.  We model the emission
region as a screen of half-wave dipole radiators resonant around the
frequencies of observation, the maximally bright emission mechanism of
nonrelativistic charges, and calculate the implied charge bunching.  From
this we infer the minimum electron energy required to overcome electrostatic
repulsion.  If FRB are the counterparts of Galactic events, their Galactic
counterparts may be detected from any direction above the horizon by radio
telescopes in their far sidelobes or by small arrays of dipoles.
\end{abstract}
\pacs{95.30.Gv,95.85.Bh,98.70.-f} \maketitle
\section{Introduction}
Lorimer, {\it et al.\/}~\cite{L07} discovered a fast (intrinsic duration
$< 5\,$ms) radio burst (FRB) in a band 300 MHz wide around $\nu = 1400\,$MHz
with a chirp indicating a dispersion measure ${\rm DM} = 375\,$pc/cm$^3$ and
a fluence ${\cal F}_\nu \approx 150\,$Jy-ms.  This dispersion measure is
consistent with propagation through the intergalactic medium from redshift
$z = 0.3$ and inexplicable as the result of Galactic plasma, but it is not
possible to constrain the contribution of plasma local to the emitter.

Thornton, {\it et al.\/}~\cite{T13} discovered four FRB in a band about 400
MHz wide around $\nu = 1400\,$MHz, with intrinsic durations $\lesssim 1\,$ms
(one burst, like that of \cite{L07}, was temporally resolved, but their
widths are explained as multipath dispersion of travel times).  Their
measured fluences ${\cal F}_\nu$ were between 0.6 Jy-ms and 8.0 Jy-ms.
Observed chirps are explicable as dispersion by intergalactic plasma,
indicating $0.5 \lesssim z \lesssim 1.0$.  The total energy radiated in the
band of observation, assuming isotropy, was (for the most luminous burst)
only about $10^{40}$ ergs, and the corresponding lower bound on luminosity
was $10^{43}$ ergs/s.  These energies and powers can be provided by a wide
range of processes involving compact objects.  The upper bound on duration
may be a more significant constraint, but is consistent with the light
travel time across neutron stars and stellar mass black holes.

The purpose of this paper is to consider the inferences that can be drawn
directly on physical grounds from the observed FRB phenomenology.  Unlike 
\cite{FR13,PP13,TT13,KIM13,LSM13}, it is not to develop an astronomical
model or to identify source objects.

As was realized long ago for radio pulsars \cite{LS68}, such intense
emission from a small source, implied by its short duration, corresponds
to a brightness temperature $T_b$ far in excess of any possible equilibrium
temperature or even particle energy.  A radiation field at a specified
frequency interacts with a limited range of particle momenta $p$.  If the
particles are uncorrelated and their distribution function $f(p)$ in that
range is fitted to an equilibrium distribution at temperature $T_{part}$,
then $T_b \le T_{part}$ \cite{PS65}.  For a relativistic power law $f(p)
\propto p^{-\alpha}$ with $\alpha > -2$
\begin{equation}
k_B T_b \le k_B T_{part} = {pc \over \alpha + 2};
\label{Tb}
\end{equation}
in general, $k_B T_{part} \simeq pc$.  Because a power law distribution is
nonequilibrium, thermodynamics permits arbitrarily high $T_b$.  However,
unless there is a population inversion (an unprecedented $\alpha < -2$,
implying $T_{part} < 0$, in which case there is no bound on $T_b$) or
coherent emission, $T_b$ is limited by Eq.~\ref{Tb}.
%  If the radiation were emitted by uncorrelated particles with
%a narrow range of momenta and with a distribution function in that range
%well fitted by an equilibrium distribution at finite temperature, the
%brightness temperature of the radiation could not exceed that temperature 
%\cite{PS65}.  In general, even for a power law distribution, the locally
%fitted temperature is comparable to the particle energy.  This applies even
%if the distribution function is nonequilibrium because the radiation field
%at any specified frequency is coupled only to a portion of it locally
%described by a temperature; nonequilibrium thermodynamics would permit an
%arbitrary brightness temperature, but radiation kinetics does not.
%Only if there is a population inversion, which would require an 
%unprecedented $\alpha < -2$, can $k_B T_b$ much exceed the energy of the
%emitting particles.

A high brightness temperature requries coherent emission by correlated
``bunches'' of particles \cite{LS68}.  Exponential amplification of a
radiation field by an inverted particle distribution function is one
process by which particles may be bunched and radiate coherently.  Plasma
instabilities are another such process, in which bunching is produced by
charged particles interacting with each other by near-zone, rather than
radiation, fields.

If emission is produced by bunches of charge $q$
with a power law momentum distribution the bunches may be regarded as
quasi-particles.  For such a nonequilibrium particle distribution function,
$p$ in Eq.~\ref{Tb} is replaced by the momentum of the bunch, $qp/e$:
\begin{equation}
k_B T_b \le k_B T_{bunch} = {q \over e}{pc \over \alpha + 2}. 
\end{equation}
This upper limit can be approached if coherent bunches survive for the time
required for them to equilibrate with the radiation field.

The frequency structure of ${\cal F}_\nu$ in FRB 110220, comprising bands
approximately 100 MHz wide ($\Delta \nu/\nu \approx 0.1$ \cite{T13}), is an
important clue.  It is evidence for the spatial structure of coherent
emission, perhaps as the consequence of a collective interaction (plasma
instability \cite{W98,ZZZ12}); the incoherent emission of randomly
distributed charges would not show such frequency structure.  This frequency
structure is also inconsistent with radiation by relativistically moving 
charges or bunches (synchrotron or curvature radiation) because that
produces a broad-band $\Delta \nu/\nu \approx 1$ spectrum \cite{J98}, even
if they are monoenergetic.  For this reason we consider radiation by
particles moving nonrelativistically in the source frame.  However, the
source frame may be moving towards us with a Lorentz factor $\Gamma \gg 1$,
in analogy to a gamma-ray burst (Thornton, {\it et al.\/} \cite{T13} argued
against observed GRB as sources of FRB on the basis of their event rates and
the absence of associations with the observed FRB).
\section{The fast radio bursts}
Here we apply the brightness temperature argument to the most intense burst,
FRB 110220, for which ${\cal F}_\nu = 8.0$ Jy-ms and $z = 0.81$.  With only
an upper bound to the FRB duration, this argument can only set limits, so we
ignore an order-of-unity error and take a static Newtonian universe with the
source at a distance $D = 10^{28}$ cm (3 Gpc).
%Following conclusion invalid, based on erroneous assumptions that clumping
%factor equals ratio of T_b to particle energy, assumption of curvature
%radiation with radius of curvature equal to source size and peak frequency
%equal to frequency of observation
%For a quasi-static source, such as a neutron star magnetosphere, the
%degree of clumping is so large that the particles' Coulomb repulsion is
%inconsistent with their energies as implied by the observed frequency of
%emission.  The brightness temperature constraint is mitigated if the source
%region is expanding relativistically, and we find a lower bound on the
%Lorentz factor $\Gamma$ of expansion.
For a source of (unmeasured) duration $\Delta t$ but measured (unpolarized)
fluence spectral density ${\cal F}_\nu$, the flux density is
\begin{equation}
F_\nu \approx {{\cal F}_\nu \over \Delta t}{D^2 \over \Delta x^2},
\end{equation}
where $\Delta x$ is the size of the region illuminating the observer and we
have assumed isotropic emission at the source.  For a static source $\Delta
x$ is its geometrical size, but for a relativistically expanding source
\begin{equation}
\label{deltax}
\Delta x \approx c \Delta t \Gamma,
\end{equation}
where the factor of $\Gamma$ comes from the relativistic beaming of the
radiation emitted from a shell of radius $R \approx c \Delta t \Gamma^2$
into an angle $\approx 1/\Gamma$.  The brightness temperature in the
observer's frame is 
\begin{equation}
k_B T_{b,obs} \equiv {1 \over 2} {F_\nu c^2 \over \nu^2} \approx {1 \over
2} {{\cal F}_\nu \over \Delta t}{D^2 \over \Delta x^2} {c^2 \over \nu^2}.
\end{equation}
Transforming to the source frame, using the scalings ${\cal F}_\nu \propto
\Gamma^0$ (because the bandwidth scales with the frequency), $\Delta t
\propto \Gamma$, $\nu \propto \Gamma^{-1}$ and substituting (\ref{deltax})
\begin{equation}
k_B T_{b,src} \approx {1 \over 2} {{\cal F}_\nu \over \nu^2} {D^2 \over
\Delta t^3} {1 \over \Gamma} \approx {2 \times 10^{21} \over \Gamma} 
\Delta t_{-3}^{-3} \ {\rm ergs},
\end{equation}
where $\nu$, ${\cal F}_\nu$ and $\Delta t$ are the observed quantities and
$\Delta t_{-3} \equiv \Delta t/(1\,{\rm ms})$, yields $T_{b,src} > 
10^{37}/\Gamma^{\,\circ}$K!  It is evident that the radiation must be
coherent because particles cannot be accelerated to energies ${\cal O}
(k_B T_{b,src})$ for any possible $\Gamma$.  Even if the sources were within
the Galactic disc ($D \simeq 100$ pc), the lower bound on $T_{b,src}$ would
imply coherent emission.

%Because the observed
%radiation is at a comparatively low frequency (compared to the frequencies
%associated with very energetic particles), and because momentum transverse
%to a magnetic field is rapidly radiated away and not replenished by parallel
%electric fields, we assume a radiation process that minimizes the radiation
%frequency, curvature radiation with a radius of curvature $R$.
%Preceding probably not valid assumption.  Leads to \gamma = 1/(\nu \Delta t)
%> 10^6.  Could have smaller radius of curvature, lower \gamma.
If the radiation is powered by the dissipation of magnetic energy, we can
set a lower bound on the magnetic field:
\begin{equation}
B^2 > {8 \pi \over c} {{\cal F}_\nu \Delta \nu \over \Delta t} {D^2 \over
\Delta x^2} \approx {8 \pi {\cal F}_\nu \Delta \nu  D^2 \over c^3 \Delta t^3
\Gamma^3} \approx {10^{19} \over \Gamma^3 \Delta t_{-3}^3}\ {\rm gauss}^2.
\end{equation}
This suggests magnetic reconnection of neutron star fields, or of white
dwarf fields if $\Gamma \gg 1$.  The energy flux $B^2 c/8\pi$ is consistent
with the upper bound ${\cal O}(10^{29}$ erg/cm$^2$-s) on the power
density in nonthermal particles set by cascading thermalization into opaque
equilibrium pair plasma at higher energy density \cite{K96,K97,K06}.

There are similarities between the FRB and the nanosecond ``nanoshots'' of
the Crab pulsar \cite{HKWE03,HE07,Ly07,L08}, even though the energy scales
differ by a factor ${\cal O}(10^{12})$.  The inferred brightness
temperatures are of similar orders of magnitude, although this is only a
very rough comparison because of the likelihood of relativistic expansion
(at unknown and different Lorentz factors) towards the observer.  More
significant is the similarity in spectral structure: FRB and nanoshots both
display bands of width $\Delta \nu/\nu \approx 0.1$ \cite{T13,HE07}.  If
this width is interpreted as radiation damping, it suggests radiation by
impulsively excited oscillations with $Q \approx 10$.  In both classes of
source the spectral structure may instead be interpreted as harmonic
emission (with harmonic index ${\cal O}(10)$) of a fundamental frequency
$\Delta \nu$, perhaps close to the electron plasma frequency of a strongly
turbulent plasma \cite{W98}.
\section{Dipole emission model}
Radiation by nonrelativistically moving particles may be treated by a
multipole expansion, with the dipole term generally dominant \cite{J98}.
Following the argument in the Introduction, we suggest that it is useful
to consider the hypothesis that the emission mechanism in FRB may be
described by the coherent emission of non-relativistically moving (in the
source frame) clumps of charge.

We model the emission region as a surface covered with half-wave dipole
antennas at the observed frequency, and estimate the charge $q$ that must
flow in each in order to produce the observed brightness temperature. 
This is a minimal model of radiation by nonrelativistic accelerated charges
\cite{J98}.  Half-wave (length $L = \lambda/2$, where $\lambda$ is the
wavelength) dipoles are nearly maximally efficient emitters, and lead to the
least restrictive demands on the bunching of charges.  Because these
antennas are approximately impedance-matched to free space they are
effective absorbers as well as emitters, so that radiation emitted behind
this surface screen is absorbed and is not observed.

The dipoles are not meant as a physical model, but only as a representation
of the coupling between source and radiation field that may be applied to
generic nonrelativistic radiation mechanisms, not limited to coherent
electron plasma wave turbulence \cite{UK00}.  The impedance of an ideal
$\lambda/2$ dipole $(73 + 42.5 i)\,\Omega$ is close enough to that of free
space ($377\,\Omega$) that radiating structures of approximately that
dimension may have the inferred $Q \approx 10$.  Structure on other length 
scales (in units of the radiated wavelength) radiate inefficiently; it may
be present, but almost all the radiation is produced by structure on the
scale of $\lambda/2$.  For a broad-spectrum source, the radiation at any
wavelength $\lambda$ is produced by structure (effectively dipoles) with
$L \approx \lambda/2$ or spatial structure factor $k \approx \pi/\lambda$.
The screen of $\lambda/2$ dipoles is a fair approximation to many turbulent
radiation sources.

A single dipole with oscillating charge $q$ and dipole moment $q \lambda/2$
radiates a power
\begin{equation}
P_{dipole} \approx {4 \pi^4 \over 3} {\nu^2 q^2 \over c}.
\end{equation}
We assume that the dipoles are not identical, but that their oscillation
frequencies are spread over a bandwidth $\Delta \nu \approx \nu$.  A sphere
of radius $R = c \Delta t \Gamma^2$ is covered by approximately $16 \pi R^2
/\lambda^2$ dipoles.  Equating the total radiated power to the observed
power $4 \pi D^2 {\cal F}_\nu \nu/\Delta t$ yields
\begin{equation}
q^2 \approx  {3 \over 16 \pi^4} {{\cal F}_\nu D^2 c \over (\nu \Delta t)^3
\Gamma^4}.
\end{equation}
The bunching factor
\begin{equation}
\label{qovere}
{q \over e} \approx 2.7 \times 10^{19} \Gamma^{-2} \Delta t_{-3}^{-3/2}.
\end{equation}
This result applies to both isotropic and beamed emission, the latter
possible if the dipoles are appropriately phased, as might be the case if,
for example, the radiating elements are charge bunches in relativistic
motion.  Although the dipoles are only weakly coupled and radiate
approximately independently, the fact that a single burst is observed
indicates that they are excited by a common larger scale event.

The total number of electrons radiating, assuming isotropic emission, is
\begin{equation}
N_e = 4 \pi {R^2 \over (\lambda/2)^2}{q \over e} = {4 \sqrt{3} \over \pi}
{D^2 \Gamma^2 \sqrt{{\cal F}_\nu c \nu \Delta t} \over e} \approx 2.7 \times
10^{33} \Gamma^2 \Delta t_{-3}^{1/2}.
\end{equation}
The mass of neutralizing protons is only $4 \times 10^9 \Gamma^2 \Delta 
t_{-3}^{1/2}$ g; alternatively, the radiating plasma may be a pair gas
without baryons.  The potential associated with the charge $q$ is $V
\approx 2 q / \lambda$ and the electrostatic energy per electron
\begin{equation}
\label{eV}
eV \approx {2 e q \over \lambda} \gtrsim 4 \times 10^{11} \Gamma^{-2}
\Delta t_{-3}^{-3/2}\ {\rm eV}.
\end{equation}
This implies a minimum electron Lorentz factor to permit bunching
\begin{equation}
\label{gamma}
\gamma = eV /m_ec^2 \gtrsim 10^6/(\Gamma^2 \Delta t_{-3}^{3/2}).
\end{equation}
Even $\gamma \gg 1$ need not invalidate the description of the radiation as
that of a screen of half-wave dipoles of charge moving nonrelativistically,
provided the phase velocities of the coherent charge bunches are
nonrelativistic.  This is consistent with relativistic energies of
individual electrons (a relativistic plasma may support waves with
nonrelativistic phase velocities, or bunches may be confined magnetically).
A similar argument applies to the nanoshots of the Crab pulsar.

If, however, $\gamma \simeq 1$, then Eq.~\ref{gamma} implies
\begin{equation}
\label{Gamma}
\Gamma \gtrsim 1000 \Delta t_{-3}^{-3/4}.
\end{equation}
This is larger than values of $\Gamma$ inferred for gamma-ray bursts, but
perhaps by less than an order of magnitude, hinting at but not requiring
related processes.

The electrostatic energy in the electrons
\begin{equation}
E_e = N_e e V = {6 \over \pi^3}{{\cal F}_\nu D^2 \over \Delta t} \approx
1.5 \times 10^{33} \Delta t_{-3}^{-1}\,{\rm erg}.
\end{equation}
This is about $10^{-7}$ of the energy implied by the observed fluence.  The
energy of the radiating electrons must be replenished (for example, by an
ongoing plasma instability), or the electrons themselves replaced by others
equally energetic and bunched, in about $10^{-7}$ of the burst duration, a
time $< 10^{-10}$ s.

The bunching factor required to explain the inferred $T_{b,src}$ as
radiation by a power-law distribution of quasi-particles may be compared to
that required for $\lambda/2$ dipole emission.  For a power-law distribution
of quasiparticle momenta $q/e \approx 2 \times 10^{27} (m_e c/p) \Delta
t_{-3}^{-3} \Gamma^{-1}$, taking $p$ to be the value required to explain the
observed momentum, which depends on unknown parameters such as the magnetic
radius of curvature.  This $q/e$ can be much greater than that of
Eq.~\ref{qovere}, as would also be the corresponding electrostatic energy.
\section{Discussion}
The short observed durations of FRB imply either a very small source region
or relativistic expansion; $R \approx 3 \times 10^7 \Delta t_{-3}
\Gamma^2\,$cm.  In the latter case (\ref{Gamma}) leads to the estimate $R
\approx 3 \times 10^{13}\,$cm, with no obvious astronomical identification.
However, $\Gamma$ and $R$ may be much smaller, provided $\gamma$ is large
(\ref{gamma}).  This might describe a source confined within a static
magnetosphere, for which $\Gamma = 1$ \cite{PP07}.

Predictions \cite{CJ60,CN71,UK00} of fast radio bursts do not match
the observed FRB, and subsequent explanations
\cite{FR13,PP13,TT13,KIM13,LSM13} address the event frequency and energetics
but not the coherent emission.  We cannot exclude the possibility that if
the rate of GRB-like events exceeds (because they are beamed or radiate
outside the soft gamma-ray band) the observed rate of GRB by a factor
$\gtrsim 10^3$, then FRB may be a GRB epiphenomenon.  If so, then
$d\ln{N_{FRB}}/d\ln{{\cal F}_{FRB}} \to 0$ as ${\cal F}_{FRB} \to 0$ because
their sources are discrete and finite in number; they may  be detected out
to a cosmic horizon.  On the other hand, if FRB result from sources, such as
stellar flares, that have no intrinsic scale but occur with increasing
frequency at small energies, then $d\ln{N_{FRB}}/d\ln{{\cal F}_{FRB}}$ is
determined by the geometry of the Universe and the evolution of the source
population, and is -3/2 for a non-evolving Newtonian cosmology.

The sources of FRB may also make novel fast ($\lesssim$ ms) events, as yet
unobserved, at other frequencies, but the lower sensitivity of 
quantum-limited detectors and the likely absence of coherent emission at
shorter wavelengths may preclude detection.  Clumps of net charge $q$ on
scales ${\cal O}(10\,$cm) (radiating coherently at 1400 MHz) imply maximum
interparticle distances ${\cal O}[10(e/q)^{1/3}\,{\rm cm}] \approx 300\,
\Gamma^{2/3} \Delta t_{-3}^{1/2}\,$\AA.  Spatial coherence on scales
$\ll 10\,{\rm cm}$, with correspondingly high brightness temperatures at
shorter (even visible) $\lambda$, requires a total electron density $n_e \gg
(2/\lambda)^3$.  This may be a more stringent requirement than the $n_e >
(2/21\,{\rm cm})^3 q/e$ required by the FRB.  If structure exists on small
enough scales, coherent emission in visible light is possible, but is not
implied by the radio observations.  

If the sources of FRB are found also in our Galaxy, Galactic events will
be ${\cal O}(10^{11})$ times brighter than those at cosmological distances
and observable outside a telescope's nominal beam if they are above
the horizon.  Antenna sidelobes at large ($\gtrsim 30^{\,\circ}$) angles
are not easily measured, but for the Parkes 64 m telescope used by
\cite{L07,T13} they are estimated \cite{BS11} to be suppressed compared to
the main beam by no more than 59 dB, while the far sidelobes of the off-axis
Green Bank Telescope are suppressed by 78 dB \cite{B11}.  Search of any
observing record (such as pulsar surveys \cite{BS11,BNM12}) with ms time
resolution and de-dispersing software can detect or set bounds on the
frequency of Galactic FRB.  In a multi-beam instrument such as that used in
the Parkes Multibeam Pulsar Survey out-of-beam events will occur with nearly
equal intensity in each beam (unfortunately, a characteristic shared with
terrestrial interference).  The absence of any credible event in the PMPS
\cite{BNM12}, which involved approximately 2000 hours of observing, sets an
upper bound of about 10/year on the rate of such Galactic events.

Even a single $\lambda/2$ dipole antenna has a sensitivity $\approx 0.1
(\lambda/D)^4 \approx 10^{-11}$ times that of the main beam of an aperture
of diameter $D$, where the numerical estimate applies to the Parkes
telescope at 1400 MHz.  Galactic events of the same luminosity as those
reported in-beam \cite{L07,T13} are detectable from any direction above the
horizon by arrays of a small number of dipoles, also providing rough
directional information.  They may be distinguished from terrestrial
interference by ms time resolution, processing for plasma dispersion and
requiring detection at several widely separated sites, adding long
interferometric baselines to constrain localization.   An array of 10--100
dipoles tuned to the inferred ($\approx 2000\,$MHz) source frequencies of
cosmological FRB could directly test all hypotheses, including that of
association with giant SGR outbursts \cite{PP13}, that they are produced by
Galactic events that occur at least a few times during the duration of
observations.  If the radio transient in its source frame extends to
frequencies (10--240 MHz) in the LOFAR \cite{LOFAR} band, that instrument
will either detect them or provide much tighter constraints.  
\section{Note Added}
Recent papers \cite{LG14,KONZJ14} have constrained the astronomical
environment of the sources of FRB.
 
\begin{acknowledgments}
I thank J. Goodman, C. Heiles, T. Piran and K. Postnov for useful discussions.
\end{acknowledgments}

\end{document}